# Anisotropic Spin Stripe Domains in Bilayer $La_3Ni_2O_7$

Naman K Gupta,[1,†] Rantong Gong,[1,†] Yi Wu,[2,†] Mingu Kang,[2,3,4] Christopher T. Parzyck,[2] Benjamin Z. Gregory,[2,3] Noah Costa,[1] Ronny Sutarto,[5] Suchismita Sarker,[6] Andrej Singer,[3] Darrell G. Schlom,[3,4,7] Kyle M. Shen,[2,4*] and David G. Hawthorn[1*]

[1]Department of Physics and Astronomy, University of Waterloo, Waterloo, N2L 3G1, Canada
[2]Laboratory of Atomic and Solid State Physics, Department of Physics, Cornell University, Ithaca, NY 14853, USA
[3]Department of Materials Science and Engineering, Cornell University, Ithaca, NY 14853, USA
[4]Kavli Institute at Cornell for Nanoscale Science, Cornell University, Ithaca, NY 14853, USA
[5]Canadian Light Source, Saskatoon, Saskatchewan, S7N 2V3, Canada
[6]Cornell High Energy Synchrotron Source, Cornell University, Ithaca, NY 14853, USA
[7]Leibniz-Institut für Kristallzüchtung, Max-Born-Straße 2, 12489 Berlin, Germany

*To whom correspondence should be addressed;
†These authors contributed equally to this work.
E-mail: kmshen@cornell.edu and david.hawthorn@uwaterloo.ca.

## Abstract

**The discovery of superconductivity in $La_3Ni_2O_7$ under pressure has motivated the investigation of a parent spin density wave (SDW) state which could provide the underlying pairing interaction. Here, we employ resonant soft x-ray scattering and polarimetry on thin films of bilayer $La_3Ni_2O_7$ to determine that the magnetic structure of the SDW forms unidirectional diagonal spin stripes with moments lying within the $NiO_2$ plane and perpendicular to $Q_{SDW}$, but without evidence of the strong charge disproportionation typically associated with other nickelates. These stripes form anisotropic domains with shorter correlation lengths perpendicular versus parallel to $Q_{SDW}$, revealing nanoscale rotational and translational symmetry breaking analogous to the cuprate and Fe-based superconductors, with possible Bloch-like antiferromagnetic domain walls separating orthogonal domains.**

# Introduction

The discovery of superconductivity with a transition temperature ($T_c$) above 80 K under hydrostatic pressure has ignited intense interest in $La_3Ni_2O_7$ (*1*) as a new platform to compare against the Cu- and Fe-based superconductors to understand which common ingredients are essential for realizing high $T_c$'s. (*2,3*) In both the cuprates and Fe-based families, parent antiferromagnetic states as well as nematicity are both known to play critical roles in their phase diagrams, the detailed understandings of which are essential for developing microscopic models. Presently, our understanding of the parent state from which the superconductivity condenses in $La_3Ni_2O_7$ remains incomplete. Recent resonant inelastic x-ray scattering (RIXS) (*4*), heat capacity and magnetic susceptibility (*5,6*), nuclear magnetic resonance (NMR) (*7,8*) and $\mu$SR (*9,10*) measurements of bulk $La_3Ni_2O_7$ suggest the presence of a spin density wave (SDW) transition occurring around $T_{\mathrm{SDW}} \approx 150$ K, with a wavevector of $\left(H = \frac{1}{4}, K = \frac{1}{4}\right)$ deduced from RIXS. (*4*) However, essential information regarding the microscopic spin structure and orientation remains unresolved.

Here, we present detailed resonant soft x-ray scattering (RSXS) and polarimetry measurements on epitaxial thin films of bilayer $La_3Ni_2O_7$ which reveal that the detailed spin structure consists of equal domains of bicollinear spin stripes, with the spins lying in the $NiO_2$ plane and aligned perpendicular to the SDW wavevector. Temperature-dependent spectroscopic measurements support a scenario with only a single electronic Ni site, with no evidence for charge or bond disproportionation across $T_{SDW}$, and where the oxygen ligand holes are strongly hybridized with the Ni *d* orbitals. The SDW domains exhibit a surprising anisotropy in their correlation lengths parallel versus perpendicular to their wavevector, suggesting that nanoscale rotational symmetry breaking might also play an important role in $La_3Ni_2O_7$, reminiscent of the nematic order found in the cuprates and Fe-based superconductors. (*11,12,13*) Finally, a detailed polarization and temperature-dependent analysis of the SDW Bragg peaks suggests the presence of extended, Bloch-like domain walls between orthogonal domains.

One of the challenges of investigating $La_3Ni_2O_7$ is the recent discovery of multiple structural polymorphs, including the expected $n = 2$ Ruddlesden-Popper structure comprised of bilayers of $NiO_6$ octahedra (dubbed 2222), and a surprising naturally-formed superlattice of alternating single-layer and trilayer blocks of $NiO_6$ octahedra (dubbed 1313) (*14,15*). It is plausible that both polymorphs could exist within a single macroscopic bulk crystal sample, and both structures have been reported to be the superconducting phase (*15*). Layer-by-layer reactive oxide molecular beam epitaxy (MBE) synthesis of thin films offers a solution to addressing this polymorphism. Here, we focus our investigation on phase-pure epitaxial thin films of the bilayer 2222 polymorph of $La_3Ni_2O_7$. Epitaxial thin films of 16 nm thickness were grown on $NdGaO_3$ (110) substrates using reactive oxide molecular beam epitaxy with shuttered deposition at a pressure of $8 \times 10^{-7}$ torr of 80 % distilled ozone and a temperature of 800° C. Samples were characterized by lab-based x-ray diffraction, electrical transport, and synchrotron-based hard x-ray diffraction (see Supplementary Information Section II). 6 samples were synthesized and investigated and all exhibited qualitatively similar behaviors; data from two samples are shown in this manuscript.

# Results

In Figure 1b, we show RSXS measurements on the Ni $L_3$ edge in $\pi$-polarization along the (*H, H,* 1.86) direction for a series of temperatures. Here, we employ the pseudo-tetragonal unit cell such at $a_T$ and $b_T$ parallel to the in-plane NiO bond directions. Below 160 K, a sharp peak emerges with $\mathbf{Q}_{\mathrm{SDW}}(20\ \mathrm{K}) = (0.2508, 0.2508, L)$, slightly incommensurate, but within experimental accuracy of the commensurate value $\left(\frac{1}{4}, \frac{1}{4}, L\right)$. The peak intensity is approximately 45 times stronger in $\pi$ versus $\sigma$ polarization at 20 K, suggesting that the scattering originates from magnetic rather than charge ordering, consistent with previous measurements. (*4*) The resonance energy dependence of the high temperature background subtracted scattering intensity, $I_S(\hbar\omega, 100\ \mathrm{K}) = I\,(\hbar\omega, 100\ \mathrm{K}) - I(\hbar\omega, 220\ \mathrm{K})$, at the SDW peak maximum for $L$ = 1.86 is shown in Fig. 1c, together with the x-ray absorption spectrum (XAS). This demonstrates that the $\left(\frac{1}{4}, \frac{1}{4}\right)$ Bragg peak is peaked only at the Ni $L$ resonances, and not on the La $M_4$ edge or off-resonance.

In Figure 1f, we show the $L$ dependence of the SDW structure factor, $I_L$, at the in-plane $\left(\frac{1}{4}, \frac{1}{4}\right)$ wavevector on the Ni $L_2$ edge (870 eV). Note, the $L_2$ edge is chosen for this purpose for its higher energy and thus larger accessible range of $L$ relative to the Ni $L_3$ edge. While the scattering intensity $I_S(L)$ is peaked at an incommensurate $L$ value (see Supplementary Information Section VI), assessing the $L$ dependence of the SDW structure factor, $I_L$, requires correction for the variation in the measurement geometry during the $L$ scan to account for both the absorption of the incident and scattered x-rays and the projection photon polarization onto the crystallographic axes. These corrections are akin to the well-known polarization and absorption corrections of conventional x-ray diffraction but generalized for absorption from a thin film and for resonant scattering from the in-plane magnetic structure identified below (see Supplementary Information Section VI). As shown in figure 1f, $I_L$ is peaked at $L$ = 2, but with only a 14 ± 4 Å correlation length along the $c$ - direction, revealing the highly two-dimensional nature of the magnetic order in $La_3Ni_2O_7$.

We note that both the resonance energy and $L$ dependence of $\left(\frac{1}{4}, \frac{1}{4}\right)$ SDW Bragg peak establish that the SDW intrinsically originates from $La_3Ni_2O_7$ and not an impurity phase, as has plagued the identification of ostensible charge or spin density order in other related nickelates. (*16,17*) Specifically, and the energy dependence of the scattering is consistent with magnetic order, with a similar energy dependence to other magnetically ordered nickelates (*18,19,20*), and is inconsistent with a structural superlattice, such as staged oxygen impurities. The $L$ dependence peaked an integer $L$ value indicates the SDW order corresponds to the $c$-axis lattice constant of $La_3Ni_2O_7$ as opposed to other magnetically ordered compounds such as $RENiO_3$ (*18,19,20*) or $La_2Ni_2O_5$ *(21)* that also exhibit (1/4, 1/4 , *L*) magnetic order but have different lattice constants and/or order peaked at non-integer values of *L*.

In many related nickelates in the Ruddlesden-Popper sequence, including $n$ = 1 $La_{2-x}Sr_xNiO_4$ and $n = \infty$ perovskite $RENiO_3$ (RE = Nd, Pr, Sm), SDW order coincides with strong charge or bond disproportionation, where the local electronic and orbital environment of the Ni sites is strongly modulated at an atomic scale.*(18,19,22,23,24,25,26,27)* Whether such behavior also occurs in $La_3Ni_2O_7$ remains an important open question. Evidence of charge order or bond disproportionation can be most directly identified from the observation of structural changes and superlattice charge peaks. In addition, charge order may exhibit anomalies in transport and thermodynamic properties, such the metal-insulator transition that is tied to charge disproportionation in $RENiO_3$ (*24,25*), semiconductor-insulator transitions tied to spin-charge order in $La_{2-x}Sr_xNiO_4$ (22) or $La_4Ni_3O_8$ (28,29), or the metal-metal transition tied to spin-charge order in $La_4Ni_3O_{10}$ (30). In cuprate superconductors, charge order can also be evident in a change in the Hall and Seebeck coefficients, indicative of Fermi surface reconstruction with the onset of CDW order. (31)

Evidence of bond disproportionation in $RENiO_3$ is also found in resonant x-ray diffraction and x-ray absorption spectroscopy. In $RENiO_3$, the onset of bond disproportionation leads to spectroscopically distinct Ni sites that can be identified as temperature dependent changes in the XAS across the metal-insulator transition. (32,33,34). With resonant x-ray scattering, signatures of bond disproportionation can be detected in the energy and polarization dependence of the resonant scattering. Ni sites having a different orbital occupation and local crystal field, such as long-bond and short-bond Ni sites in $RENiO_3$, will modulate the x-ray scattering form factors on the different Ni sites leading them to resonate at slightly different energies. (35) These distinct Ni sites also exhibit a different magnitude and orientation of the spin. As the scattering intensity depends on the orientation of the x-ray polarization with respect to the spin, Ni sites with distinct charge environments can result in the linear dichroism of the SDW peak intensity ($\sigma$ versus $\pi$ incident polarization) becoming energy dependent (see Supplementary Information Section III), as reported in $RENiO_3$ heterostructures. (18)

Considering these signatures of charge order, we find no evidence for charge order or charge disproportionation in our samples of $La_3Ni_2O_7$. We observe no distinct changes in the Ni *L* edge XAS spectra across $T_{SDW}$ (Figure 1e) and the linear dichroism of the $\left(\frac{1}{4},\frac{1}{4}\right)$ peak in $La_3Ni_2O_7$ does not exhibit any observable energy dependence (Figure 1d), suggesting that all Ni sites have the same orbital occupation and magnitude of spin in the SDW state. In addition, we do not detect strong anomalies in either the resistivity or Hall coefficient (See Supplementary Information Figure S1), nor do we identify evidence of structural phase transitions or superlattice peaks from hard x-ray diffraction (See Supplementary Information section II). While these measurements collectively constitute a null result, they suggest that the dominant order parameter is magnetic and that unlike many other nickelates, any associated charge modulation in our samples is either absent, does not onset near $T_{SDW}$ or is too weak to be detected in our present measurements.

# Measurement of Spin Configuration

We now determine the orientation of the staggered moments at 20 K, deep within the SDW state. For this we make use of the sensitivity of the intensity of resonant scattering to the orientation of the photon polarization relative to the magnetic moments, analogous to polarized neutron scattering. (18,26,35,36). The resonant elastic x-ray scattering cross-section is given by: (35,37)

$$I^{\mathrm{cr}}(\epsilon_{\mathrm{in}}, \hbar\omega, \mathbf{Q}) \propto \left| \boldsymbol{\epsilon}^{*}_{\mathrm{out}} \cdot \left( \sum_{j} F_j(\hbar\omega, \mathbf{Q}) e^{i\mathbf{Q}\cdot\mathbf{r}_j} \right) \cdot \epsilon_{\mathrm{in}} \right|^2 \tag{1}$$

where $\hbar\omega$ is the photon energy, $\mathbf{Q} = \mathbf{k}_{\mathrm{out}} - \mathbf{k}_{\mathrm{in}}$ is the momentum transfer and $\boldsymbol{\epsilon}_{\mathrm{in}}$ and $\boldsymbol{\epsilon}_{\mathrm{out}}$ are the incident and scattered x-ray polarization, respectively. $F_j$ is a tensor that encodes the photon energy dependence of the scattering cross-section for site $j$ in the lattice, and its elements depend on the orientation of the magnetic moment at site $j$. (35) (see Supplementary Information section III) Assuming spherical symmetry of the local valence charge density, consistent with the linear dichroism of the SDW peak being energy-independent (Fig. 1d), the scattering depends on the orientation of the staggered $\Delta\vec{m}$ as

$$\sum_j F_j(\hbar\omega, \mathbf{Q}) e^{i\mathbf{Q}\cdot\mathbf{r}_j} \propto \begin{bmatrix} 0 & \Delta m_{[001]} & -\Delta m_{[110]} \\ -\Delta m_{[001]} & 0 & \Delta m_{[-110]} \\ \Delta m_{[110]} & -\Delta m_{[-110]} & 0 \end{bmatrix}, \quad (2)$$

where $\Delta m_u$ are the components of the staggered moments in three orthogonal directions [001], [−110] and [110].

These components can then be deduced by varying the alignment of $\boldsymbol{\epsilon}_{\text{in}}$ relative to the [001], [-110] and [110] directions. In an experiment, this is achieved by rotating the sample azimuthally by an angle $\phi$ about an axis normal to the $\left[\frac{1}{4}, \frac{1}{4}, L\right]$ set of lattice planes (here $L$ is 1.93), as depicted in Figure 2a, which is accomplished by mounting the $c$-axis normal film on a 43.7° wedge. This approach enables the orientation of the moments to be rotated relative to the incident polarization, which can be set to be $\pi$, $\sigma$, circular or linear 45° ($\pi - \sigma$), with the scattering wavevector remaining centered on the $(1/4, 1/4, 1.93)$ peak. As shown in Figure 2b, the scattering intensity, measured here at 20 K, has a strong dependence on the azimuthal angle, $\phi$ as well as on the polarization of the incident light. In Figure 2c, we plot the SDW peak amplitude as a function of azimuthal angle $\phi$, for our various incident polarizations. Notably, the peak intensity exhibits a complex, non-monotonic dependence with $\phi$ and large variations with incident polarization. This azimuthal dependence can be compared to simulations of the staggered moments oriented along different directions, calculated using Equations 1 and 2, the sample geometry, and a correction for the geometry-dependent absorption of the incident and scattered x-rays (Supplementary Information Section IIIB). As shown in Figure 2c the measurements show remarkable agreement with the moments forming diagonal, bicollinear spin stripes (Figure 2d) with the magnetic moments lying entirely within the $a-b$ plane but oriented perpendicular to $\mathbf{Q}_{\text{SDW}}$. We emphasize this model was calculated without any free fitting parameters, apart from an overall scaling factor.

In contrast, the calculated polarization and $\phi$ dependence for other possible SDW scenarios (Figure 2f and g), including $\Delta\mathbf{m}$ out of the plane, $\Delta\mathbf{m} \parallel \mathbf{Q}_{\text{SDW}}$, or a non-collinear configuration, all bear no qualitative resemblance to the experimental data (Figure 2c), indicating the relative contribution of $\Delta\mathbf{m}$ out of the NiO plane or parallel to $\mathbf{Q}_{\text{SDW}}$ is less than a few percent at 20 K.

Having constrained the orientation of moments predominantly within the $NiO_2$ planes, we can now consider the full 3D magnetic structure, which is informed by the $L$ dependence of the SDW scattering cross section. The fact that the SDW order is peaked at $L = 2$ and is minimal at $L = 1.5$, (Fig. 1f) indicates that with the coupling within the bilayer (intra-bilayer coupling) and between bilayers (inter-bilayer coupling) are either both ferromagnetic or both antiferromagnetic (see Supplemental Information Section VI). Given that the reasonably straight interlayer Ni-O-Ni bond is likely to involve antiferromagnetic super-exchange and that the experimental magnon dispersion is well described by a model with a large AF bilayer coupling, (*2*) this suggests the magnetic structure is consistent with the one depicted in Figure 2e, with both inter- and intra-bilayer AF coupling.

This bicollinear double spin-stripe configuration is notably different from many other nickelates, including the non-collinear spin-spiral magnetic order in thin films of $RENiO_3$ (*18,19,20*), the collinear order with moments parallel to $\left(\frac{1}{4},\frac{1}{4},\frac{1}{4}\right)$ in ultrathin $RENiO_3$ (*18*), spin/charge stripes with moments perpendicular to the planes in the square-planar trilayer nickelate $La_4Ni_3O_8$ (*29*), the reported magnetic order, with moments along *c*, in $La_2Ni_2O_5$ (*23*), and spin-stripe order in $La_{2-x}Sr_xNiO_4$ (*38,39*), which has staggered moments oriented within the NiO planes but not perpendicular to $\mathbf{Q}_{\mathrm{SDW}}$. Staggered moments in the NiO planes and perpendicular to $\mathbf{Q}_{\mathrm{SDW}}$ are found in both single-layer $La_2NiO_{4+\delta}$ (*27*) and the trilayer $La_4Ni_3O_{10}$ (*30,40*), suggesting a potential link between the SDW in these compounds and $La_3Ni_2O_7$. However, these compounds both exhibit spin-charge stripe-order, with $La_2NiO_{4+\delta}$ being an insulator and $La_4Ni_3O_{10}$ having incommensurate spin-charge stripe order. In both wavevector and orientation of moments, the SDW order in $La_3Ni_2O_7$ is also similar to the bicollinear double spin stripe observed in FeTe (*41,42,43*).

# Anisotropic, Unidirectional Magnetic Domains

The unidirectional, stripe-like character of the antiferromagnetic order is also manifest in the shape of the SDW domains which break the rotational symmetry of the lattice. We measured the shapes and intensities of the SDW Bragg peaks in the $H-K$ plane for both sets of SDW domains with orthogonal $\mathbf{Q}$ vectors, around $\left(\frac{1}{4},\frac{1}{4},L\right)$ (in red) and $\left(\frac{1}{4},-\frac{1}{4},L\right)$ (in blue) using a two-dimensional microchannel plate detector, shown in Fig. 3a. The Bragg peaks around $\left(\frac{1}{4},\frac{1}{4},L\right)$ and $\left(\frac{1}{4},-\frac{1}{4},L\right)$ have equal intensities to within experimental uncertainty (± 3 %), indicating an equal population of domains. In both sets of domains, the SDW peaks have highly anisotropic shapes, with a 2D Lorentzian fit giving correlation lengths that are much longer parallel to $\mathbf{Q}_{\mathrm{SDW}}$ ($\xi_{\parallel} = 292$ Å) versus perpendicular to $\mathbf{Q}_{\mathrm{SDW}}$ ($\xi_{\perp} = 134$ Å).

This peak shape is consistent with anisotropic domains of unidirectional SDW order, with each domain characterized by either (1/4, 1/4) or (-1/4, 1/4) order, as depicted in Figure 3d. This anisotropy could be associated with orthorhombic structural twin domains or may be orthogonal domains of unidirectional order occurring within a single structural domain. Intriguingly, this latter scenario is reminiscent of the anisotropic charge ordering reported in underdoped $YBa_2Cu_3O_{7-\delta}$ and $Bi_2Sr_{2-x}La_xCuO_{6+\delta}$, where the CDW Bragg peaks likewise exhibit an anisotropy, with longer correlations lengths parallel to the CDW wavevector, indicative of orthogonal anisotropic unidirectional CDW domains within a single orthorhombic structural domain. (*44,45*)

# Probing Magnetic Domain walls

In Figure 2, the orientation of the staggered moment at 20 K was determined to be almost entirely within the $a-b$ plane and perpendicular to $\mathbf{Q}_{\mathrm{SDW}}$ (i.e. $\Delta\mathbf{m} \parallel [-1,1,0]$ for $\mathbf{Q}_{\mathrm{SDW}}$ = (1/4, 1/4 , *L*) domains or $\Delta\mathbf{m} \parallel [1,1,0]$ for $\mathbf{Q}_{\mathrm{SDW}}$ = (-1/4, 1/4 , *L*) domains). Insights may be gleaned from the small discrepancies between the model and measurements. In particular, measurements using $\sigma$ polarization at $\phi = 0$ are at a minimum in intensity for this magnetic orientation, with deviations of the staggered moment in the NiO planes, but parallel to $\mathbf{Q}_{\mathrm{SDW}}$ and/or along [001] required to provide a finite scattering intensity. We now leverage this sensitivity of the polarization to the staggered moment orientation to investigate how the spin configuration evolves with temperature. In Figure 4b, we show a comparison between the temperature dependence of the SDW peak when measured with $\pi$ versus $\sigma$ polarization. At low temperatures ($T < 50K$), the scattering intensity of the SDW peak measured in $\pi$ polarization, $I_\pi$, is more than an order of magnitude stronger than in $\sigma$ polarization, $I_\sigma$. Upon raising the temperature to $T_{SDW}$, $I_\pi$, smoothly decreases whereas $I_\sigma$ exhibits a non-monotonic temperature dependence, growing in intensity before peaking around 130 K, and then falling rapidly to zero at $T_{SDW}$. On the other hand, the ratio of $I_\sigma/I_\pi$, shown in figure 4c grows smoothly and monotonically with increasing temperature all the way up to $T_{SDW}$ when the peak vanishes, indicating that the component of the spins oriented away from the low $T$ configuration (Figure 2c) grows with increasing temperature, with $I_\sigma/I_\pi$ > 0.6 near $T_{SDW}$.

In addition to this anomalous temperature dependence, the width of the SDW Bragg peak is surprisingly broader when measured with $\sigma$ versus $\pi$ incident polarization (Fig. 4a and d) Note, in Fig. 4d the peak width with $\sigma$ polarization is deduced by subtracting a narrow peak resulting from ~1.5% $\pi$ incident light in the nominally $\sigma$ polarized beam, as discussed in Supplementary Information Section VII. The broader peak with $\sigma$ incident polarization indicates that the staggered moments parallel to $\mathbf{Q}_{\mathrm{SDW}}$ or along [001] (probed with $\sigma$ incident light) have a shorter correlation length at low temperature than the predominant spin configuration with staggered moments in the NiO plane and perpendicular to $\mathbf{Q}_{\mathrm{SDW}}$. This observation would be inconsistent with a uniform, temperature-dependent canting of all the spins. Instead, this could be consistent with the existence of real-space defects of the magnetic order, such as Bloch-like domain walls or antiferromagnetic skyrmion-like topological defects. Indeed, Bloch-like domain walls may naturally exist between two orthogonal unidirectional domains, where the transition between domains would necessitate a reorientation of the spins by 90$^\circ$ from one domain to its orthogonal counterpart, which could occur over an extended region, depending on the relative magnitude of the magnetic anisotropy and exchange terms.

In our interpretation of the data in terms of Bloch domain walls, the staggered moments would rotate away from [110] or [-110] and out of the $a{-}b$ plane, as depicted conceptually in Figure 5. Here, $I_\pi$ would probe the bicollinear spin stripe regions within the core of each domain, while $I_\sigma$ would be sensitive to the rotated spin component in the domain walls. The relative volume fraction of domain walls would grow with increasing temperature until SDW order is lost at $T_{SDW}$. This interpretation could explain the broader peak for $I_\sigma$ when compared to $I_\pi$, as shown in Fig. 4d. Finally, the influence of the domain walls may also be the source of the small deviations from a perfectly in-plane moment at 20 K depicted in Fig. 2c. This analysis represents a possible new approach to detecting defects in antiferromagnetic order. Unlike ferromagnetic domain walls, antiferromagnetic domain walls are often difficult to detect. (*46*) Further analysis of the polarization and temperature dependence of the (1/4 1/4) SDW peaks in $La_3Ni_2O_7$ may provide key insights into the width, density and detailed magnetic configuration of the domain walls. Such investigations may be of key importance as the magnetic configuration close to $T_{SDW}$, including a high density of domain walls, may represent the melting of long-range SDW order as the material approaches its superconducting phase.

# Discussion

A comparison between these experiments with those on bulk crystals (*4*) reveal a number of commonalities, including similar $T_{SDW}$'s, temperature dependences, photon polarization dependence, and peak widths along the ($H$ $H$ $L$) direction, suggesting that the results reported here are universal to both bulk and thin film samples.

The structure of the bicollinear spin stripe order in $La_3Ni_2O_7$ bears a strong resemblance to the magnetically ordered state in FeTe, but with a major distinction: in FeTe, the spin stripe ordering is accompanied by a monoclinic structural transition that stabilizes the magnetic order. (*43, 47, 48*) In contrast, temperature-dependent, synchrotron-based hard x-ray diffraction measurements of our $La_3Ni_2O_7$ thin films do not reveal any lowering of structural symmetry or the appearance of superlattice peaks upon entering the SDW state (see Supplementary Information Section II). Furthermore, the existence of domain walls or topological defects indicate that both orthogonal magnetic domains occur within a single $NiO_2$ plane. This may imply an inherent instability in $La_3Ni_2O_7$ to rotational symmetry breaking, possibly accompanied by disordered or frustrated nematic order, (*11*) analogous to the Fe-based and cuprate superconductors where unidirectional density wave and nematic orders are pervasive and closely intertwined with superconductivity. (*12,13*) These analogies with the cuprates and Fe-based superconductors suggest that fluctuation of this SDW state is related to the formation of the superconducting state under high pressure.

# Methods

Two samples A and B were measured for this study. A more extensive set of data was measured for sample A. However, sample B was found to have comparable peak intensity, peak width, and temperature dependence as sample A, as well as reproducing the anisotropy in the width parallel and perpendicular to $Q$ (see Supplementary Information section V).

Resonant soft x-ray scattering measurements were performed at the Canadian Light Source REIXS beamline. (*49*) The majority of the measurements were made using an energy-resolved silicon drift detector, with measurements of the peak shape utilizing a 2D micro-channel plate detector. The silicon drift detector the detector resolution corresponds to $\Delta H = 0.00124$ reciprocal lattice units, compared to a peak width of 0.0028 FWHM for $\pi$ incident polarization at 20 K (fig. 4d). This leads to an increase in the peak width due to measurement resolution of ~ 10% at base temperature with $\pi$ incident polarization. The pixel size of the 2D micro-channel plate detector used for figure 3a and b is $\Delta H = 0.00006$ reciprocal lattice units, far below the measured peak widths.

The samples were oriented using the (110) Bragg peak of the $NdGaO_3$ (NGO) substrate at 2500 eV. The in-plane orientation was performed using $\left(\frac{1}{4},\frac{1}{4},L\right)$ SDW Bragg peaks at 852 eV. The samples were found to have a $c$ axis lattice constant of 20.44 Å from lab-based x-ray diffraction. Reciprocal space maps show lattice matched to the NGO substrate, which for the [110] surface of NGO imparts in-plane lattice constants of $a_T$ = 3.844 Å and $b_T$ = 3.858 Å at 100 K. (*50*) The mosaic spread of La327 films was measured to be less than 0.007° FWHM, much smaller than the angular width of the SDW peaks.

The scattering intensities, $I_S$, shown in Figures 1b (inset), 1c, 1d, 2b, 3c and 4, are deduced by subtracting the background intensity measured above the SDW transition (at 200 K or 220 K) from the measured intensity at lower temperature.

The correlation length, $\xi_u$, along different directions $u$, was determined from Lorentzian fits to the data of the form given in eq. (3) (or using an equivalent expression in reduced co-ordinates $H$ and $K$):

$$\frac{Amplitude}{\left(\frac{(\mathbf{Q}-\mathbf{Q}_{\mathrm{SDW}})\cdot\frac{1}{\sqrt{2}}(1,1,0)}{\frac{\Gamma_{[110]}}{2}}\right)^2+\left(\frac{(\mathbf{Q}-\mathbf{Q}_{\mathrm{SDW}})\cdot\frac{1}{\sqrt{2}}(1,-1,0)}{\frac{\Gamma_{[1-10]}}{2}}\right)^2+1} \quad (3)$$

where $\mathbf{Q}=\left(\frac{2\pi}{a_T}H,\frac{2\pi}{a_T}K\right)$, $\mathbf{Q}_{\mathrm{SDW}}$ is the in-plane wavevector of the SDW peak maximum and $\Gamma_u$ is an inverse correlation length. The correlation length is given by $\xi_u = {}^{2}/_{\Gamma_u}$.

For measurements with $\sigma$ incident polarization, shown in Figure 4, the incident beam includes a small $\simeq$ 1.5% $\pi$ contribution due to synchrotron light from bending chicane magnets that is in addition to the primary $\sigma$ polarized light from the EPU. In 4d, the full width at half maximum (FWHM) with $\sigma$ polarization is determined by fitting the data in Fig.4a to a narrow peak from $\pi$ polarized light, with FWHM equal to that of a pure $\pi$ polarization (blue curve in Fig. 4d), as well as broader peak with $\sigma$ polarization (red curve in Fig. 4d).

Figure 1(a) with the crystal structure of $La_3Ni_2O_7$ was generated using VESTA (51).

# Data availability

All data needed to evaluate the conclusions in the paper are present in the manuscript and supplementary information. Correspondence and requests for materials should be addressed to K. M. Shen (kmshen@cornell.edu) and D. G. Hawthorn (david.hawthorn@uwaterloo.ca).

# References

1. Sun, H. *et al.* Signatures of superconductivity near 80 K in a nickelate under high pressure. *Nature* **621**, 493–498 (2023). URL https://www.nature.com/articles/s41586-023-06408-7.

2. Keimer, B., S. A. Kivelson, S. A., Norman, M. R., Uchida S. & Zaanen, J. From quantum matter to high-temperature superconductivity in copper oxides. *Nature* **518**, 179 (2015). URL https://doi.org/10.1038/nature14165.

3. Fernandes, R. M., Coldea, A. I., Ding, H., Fisher, R. I., Hirschfeld, P. J. & Kotliar, G. Iron pnictides and chalcogenides: a new paradigm for superconductivity. *Nature* **601**, 35 (2022). URL https://doi.org/10.1038/s41586-021-04073-2.

4. Chen, X., Choi, J., Jiang, Z. *et al*. Electronic and magnetic excitations in $La_3Ni_2O_7$. *Nat. Commun.* **15**, 9597 (2024). URL https://doi.org/10.1038/s41467-024-53863-5.

5. Liu, Z. *et al.* Evidence for charge and spin density waves in single crystals of $La_3Ni_2O_7$ and $La_3Ni_2O_6$. *Sci. China Phys. Mech. Astron.* **66**, 217411 (2022). URL https://doi.org/10.1007/s11433-022-1962-4.

6. Wu, G., Neumeier, J. J. & Hundley, M. F. Magnetic susceptibility, heat capacity, and pressure dependence of the electrical resistivity of $La_3Ni_2O_7$ and $La_4Ni_3O_{10}$. *Phys. Rev. B* **63**, 245120 (2001). URL https://link.aps.org/doi/10.1103/PhysRevB.63.245120.

7. Kakoi, M. *et al.* Multiband metallic ground state in multilayered nickelates $La_3Ni_2O_7$ and $La_4Ni_3O_{10}$ probed by $^{139}$La-NMR at ambient pressure. *J. Phys. Soc. Jpn.* **93**, 053702 (2024). URL https://journals.jps.jp/doi/10.7566/JPSJ.93.053702.

8. Dan, Z. *et al.* Pressure-enhanced spin-density-wave transition in double-layer nickelate $La_3Ni_2O_7$. *Science Bulletin* (2025). URL https://doi.org/10.1016/j.scib.2025.02.019.

9. Chen, K. *et al.* Evidence of spin density waves in $La_3Ni_2O_{7-\delta}$. *Phys. Rev. Lett.* **132**, 256503 (2024). URL https://link.aps.org/doi/10.1103/PhysRevLett.132.256503.

10. Khasanov, R. et al. Pressure-induced splitting of density wave transitions in $La_3Ni_2O_{7-\delta}$. *Nat. Phys.* (2025). URL https://doi.org/10.1038/s41567-024-02754-z.

11. Böhmer, A. E., Chu, J.-H., Lederer, S. & Yi, M. Nematicity and nematic fluctuations in iron-based superconductors. *Nat. Phys.* **18**, 1412–1419 (2022). URL https://www.nature.com/articles/s41567-022-01833-3.

12. Fradkin, E., Kivelson, S. A. & Tranquada, J. M. Colloquium: Theory of intertwined orders in high temperature superconductors. *Rev. Mod. Phys.* **87**, 457–482 (2015). URL https://link.aps.org/doi/10.1103/RevModPhys.87.457.

13. Fernandes, R. M., Orth, P. P. & Schmalian, J. Intertwined vestigial order in quantum materials: Nematicity and beyond. *Annu. Rev. Condens. Matter Phys.* 10, 133–154 (2019). URL https://www.annualreviews.org/content/journals/10.1146/annurevconmatphys-031218-013200.

14. Chen, X. *et al.* Polymorphism in Ruddlesden-Popper $La_3Ni_2O_7$: Discovery of a hidden phase with distinctive layer stacking. *J. Am. Chem. Soc.* **146**, 3640– 242 3645 (2024). URL http://arxiv.org/abs/2312.06081.

15. Puphal, P. *et al.* Unconventional crystal structure of the high-pressure superconductor $La_3Ni_2O_7$. Phys. Rev. Lett. **133**, 146002 (2024) URL https://doi.org/10.1103/PhysRevLett.133.146002.

16. Parzyck, C. T. *et al.* Absence of $3a_0$ charge density wave order in the infinite-layer nickelate $NdNiO_2$. *Nat. Mater.* **23**, 486–491 (2024). URL https://www.nature.com/articles/s41563-024-01797-0.

17. Wang, B.-X. *et al.* Antiferromagnetic defect structure in $LaNiO_{3-\delta}$ single crystals. *Phys. Rev. Mater.* **2**, 064404 (2018). URL https://link.aps.org/doi/10.1103/PhysRevMaterials.2.064404.

18. Hepting, M. *et al.* Complex magnetic order in nickelate slabs. *Nat. Phys.* **14**, 1097–1102 (2018). URL https://doi.org/10.1038/s41567-018-0218-5.

19. Scagnoli, V. *et al.* Role of magnetic and orbital ordering at the metal insulator transition in $NdNiO_3$. *Phys. Rev. B* **73**, 100409 (2006). URL https://link.aps.org/doi/10.1103/PhysRevB.73.100409.

20. Frano, A. *et al.* Orbital control of noncollinear magnetic order in nickel oxide heterostructures. *Phys. Rev. Lett.* **111**, 106804 (2013). URL https://link.aps.org/doi/10.1103/PhysRevLett.111.106804.

21. Alonso, J. A., Martínez-Lope, M. J., García-Muñoz, J. L. & Fernández-Díaz, M. T. A structural and magnetic study of the defect perovskite from high-resolution neutron diffraction data. *J. Phys.: Condens. Matter* **9**, 6417 (1997). URL https://dx.doi.org/10.1088/0953-8984/9/30/010.

22. García-Muñoz, J. L., Rodríguez-Carvajal, J. & Lacorre, P. Neutron-diffraction study of the magnetic ordering in the insulating regime of the perovskites $RNiO_3$ (R=Pr and Nd). *Phys. Rev. B* **50**, 978 (1994) URL https://doi.org/10.1103/PhysRevB.50.978.

23. Imada, M., Fujimori, A. & Tokura, Y. Metal-insulator transitions. *Rev. Mod. Phys.* **70**, 1039 (1998). URL https://doi.org/10.1103/RevModPhys.70.1039.

24. Hepting, M., Minola, M., Frano, A., Cristiani, G., Logvenov, G., Schierle, E., Wu, M., Bluschke, M., Weschke, E., Habermeier, H.-U., Benckiser, E., Le Tacon, M., & Keimer, B. Tunable Charge and Spin Order in $PrNiO_3$ Thin Films and Superlattices. *Phys. Rev. Lett.* **113**, 227206 (2014). URL https://doi.org/10.1103/PhysRevLett.113.227206.

25. Alonso, J. A., García-Muñoz, J. L., Fernández-Díaz, M. T. , Aranda, M. A. G., Martínez-Lope, M. J. & Casais, M. T. Charge Disproportionation in $R$NiO$_3$ Perovskites: Simultaneous Metal-Insulator and Structural Transition in $YNiO_3$. *Phys. Rev. Lett.* **82**, 3871 (1999). URL https://doi.org/10.1103/PhysRevLett.82.3871.

26. Scagnoli, V. *et al.* Induced noncollinear magnetic order of $Nd^{3+}$ in $NdNiO_3$ observed by resonant soft x-ray diffraction. *Phys. Rev. B* **77**, 115138 (2008). URL https://link.aps.org/doi/10.1103/PhysRevB.77.115138.

27. Tranquada, J. M., Lorenzo, J. E., Buttrey, D. J. & Sachan, V. Cooperative ordering of holes and spins in $La_2NiO_{4.125}$. *Phys. Rev. B* **52**, 3581 (1995). URL https://doi.org/10.1103/PhysRevB.52.3581.

28. Zhang, J., Chen, Y., Phelan, D., Zheng, H., Norman, M. R. & Mitchell, J. F., Stacked charge stripes in the quasi-2D trilayer nickelate $La_4Ni_3O_8$, *Proc. Natl. Acad. Sci. U.S.A.* **113**, 8945 (2016). URL https://doi.org/10.1073/pnas.1606637113.

29. Zhang, J., Pajerowski, D. M., Botana, A. S., Zheng, H, Harriger, L., Rodriguez-Rivera, J., Ruff, J. P. C., Schreiber, N. J., Wang, B., Chen, Y.-S., Chen, W. C.. Norman, M. R., Rosenkranz, S., Mitchell, J. F. & Phelan, D. Spin Stripe Order in a Square Planar Trilayer Nickelate. *Phys. Rev. Lett.* **122**, 247201 (2019). URL https://doi.org/10.1103/PhysRevLett.122.247201.

30. Zhang, J., Phelan, D., Botana, A.S. et al. Intertwined density waves in a metallic nickelate. *Nat Commun* **11**, 6003 (2020). URL https://doi.org/10.1038/s41467-020-19836-0.

31. Taillefer, L., Fermi surface reconstruction in high-*Tc* superconductors. *J. Phys.: Condens. Matter* **21**, 164212 (2009). URL https://dx.doi.org/10.1088/0953-8984/21/16/164212.

32. Piamonteze, C. *et al.* Spin-orbit-induced mixed-spin ground state in $RNiO_3$ perovskites probed by x-ray absorption spectroscopy: Insight into the metal-to-insulator transition. *Phys. Rev. B* **71**, 020406 (2005). URL https://link.aps.org/doi/10.1103/PhysRevB.71.020406.

33. Liu, J. *et al.* Strain-mediated metal-insulator transition in epitaxial ultra thin films of $NdNiO_3$. *Appl. Phys. Lett.* **96**, 233110 (2010). URL https://doi.org/10.1063/1.3451462.

34. Bruno, F. Y. *et al.* Probing the metal-insulator transition in nickelates using soft x ray absorption spectroscopy. *Appl. Phys. Lett.* **104**, 021920 (2014). URL https://doi.org/10.1063/1.4861132.

35. Haverkort, M. W., Hollmann, N., Krug, I. P. & Tanaka, A. Symmetry analysis of magneto-optical effects: The case of x-ray diffraction and x-ray absorption at the transition metal $L_{2,3}$ edge. *Phys. Rev. B* **82**, 094403 (2010). URL https://link.aps.org/doi/10.1103/PhysRevB.82.094403.

36. Hannon, J. P., Trammell, G. T., Blume, M. & Gibbs, D. X-ray resonance exchange scattering. *Phys. Rev. Lett.* **61**, 1245–1248 (1988). URL https://link.aps.org/doi/10.1103/PhysRevLett.61.1245.

37. Fink, J., Schierle, E., Weschke, E. & Geck, J. Resonant elastic soft x-ray scattering. *Rep. Prog. Phys.* **76**, 056502 (2013). URL https://dx.doi.org/10.1088/00344885/76/5/056502.

38. Lee, S.-H., Cheong, S.-W., Yamada, K., & Majkrzak, C. F. Charge and canted spin order in $La_{2-x}Sr_xNiO_4$ ($x$=0.275 and 1/3). *Phys. Rev. B* **63**, 060405(R) (2001). URL https://link.aps.org/doi/10.1103/PhysRevB.63.060405.

39. Merritt, A. M., Reznik, D., Garlea, V. O., Gu, G. D. & Tranquada, J. M. Nature and impact of stripe freezing in $La_{1.67}Sr_{0.33}NiO_4$. *Phys. Rev. B* **100**, 195122 (2019) URL https://link.aps.org/doi/10.1103/PhysRevB.100.195122

40. Samarakoon, A. M., Strempfer, J., Zhang, J., Ye, F., Qiu, Y., Kim, J.-W., Zheng, H., Rosenkranz, S., Norman, M. R., Mitchell, J. F. & Phelan, D. Bootstrapped Dimensional Crossover of a Spin Density Wave. *Phys. Rev. X* **13**, 041018 (2023). URL https://link.aps.org/doi/10.1103/PhysRevX.13.041018

41. Dai, P. Antiferromagnetic order and spin dynamics in iron-based superconductors. *Rev. Mod. Phys* **87**, 855–896 (2015). URL https://link.aps.org/doi/10.1103/RevModPhys.87.855.

42. Rodriguez, E. E. *et al.* Magnetic-crystallographic phase diagram of the superconducting parent compound $Fe_{1+x}Te$. *Phys. Rev. B* **84**, 064403 (2011). URL https://link.aps.org/doi/10.1103/PhysRevB.84.064403.

43. Li, S. *et al.* First-order magnetic and structural phase transitions in $Fe_{1+y}Se_xTe_{1-x}$. *Phys. Rev. B* **79**, 054503 (2009). URL https://link.aps.org/doi/10.1103/PhysRevB.79.054503.

44. Comin, R. *et al.* Broken translational and rotational symmetry via charge stripe order in underdoped $YBa_2Cu_3O_{6+y}$. *Science* 347, 1335–1339 (2015). URL https://www.science.org/doi/10.1126/science.1258399.

45. Choi, J. *et al.* Universal stripe symmetry of short-range charge density waves in cuprate superconductors. *Adv. Mater.* **36**, 2307515 (2024). URL https://onlinelibrary.wiley.com/doi/abs/10.1002/adma.202307515.

46. Cheong, S.-W., Fiebig, M., Wu, W., Chapon, L. & Kiryukhin, V. Seeing is believing: visualization of antiferromagnetic domains. *NPJ Quantum Mater.* **5**, 1–10 (2020). URL https://www.nature.com/articles/s41535-019-0204-x.

47. Bao, W. *et al.* Tunable $(\delta\pi, \delta\pi)$-type antiferromagnetic order in $\alpha$ Fe(Te,Se) superconductors. *Phys. Rev. Lett.* **102**, 247001 (2009). URL https://link.aps.org/doi/10.1103/PhysRevLett.102.247001.

48. Rodriguez, E. E. *et al.* Magnetic and structural properties near the lifshitz point in $Fe_{1+x}Te$. *Phys. Rev. B* **88**, 165110 (2013). URL https://link.aps.org/doi/10.1103/PhysRevB.88.165110.

49. Hawthorn, D. G. *et al.* An in-vacuum diffractometer for resonant elastic soft x-ray scattering. *Rev. Sci. Instrum.* **82**, 073104 (2011). URL https://doi.org/10.1063/1.3607438.

50. Vasylechko, L., Akselrud, L., Morgenroth, W., Bismayer, U., Matkovskii, A. & Savytskii, D. The crystal structure of $NdGaO_3$ at 100 K and 293 K based on synchrotron data. *J. Alloys Compd.* **297**, 46 (2000). URL https://doi.org/10.1016/S0925-8388(99)00603-9

51. K. Momma and F. Izumi, VESTA 3 for three-dimensional visualization of crystal, volumetric and morphology data, *J. Appl. Crystallogr.*, **44**, 1272-1276 (2011). URL https://doi.org/10.1107/S0021889811038970

# Acknowledgments

This work was primarily supported by the National Science Foundation through Grant No. DMR-2104427 (K.M.S.), the Air Force Office of Scientific Research (Grant No. FA955021−1-0168, FA9550-23−1-0161; K.M.S.) and the Natural Sciences and Engineering Research Council (NSERC) of Canada (D.G.H.). The Platform for the Accelerated Realization, Analysis and Discovery of Interface Materials (PARADIM) under Cooperative Agreement No. DMR-2039380 supported optimization of film growth parameters and structural characterization (D.G.S. and K.M.S.). The U.S. Department of Energy, Office of Basic Energy Sciences under contract no. DE-SC0019414 (B.Z.G, A.S., D.G.S., K.M.S.) supported the hard x-ray synchrotron diffraction work at CHEXS. Additional support for materials synthesis was provided by the Gordon and Betty Moore Foundation's EPiQS Initiative through Grant Nos. GBMF3850 and GBMF9073 (D.G.S.). N.K.G. acknowledges support from the Waterloo Institute of Nanotechnology (WIN). This work is based on research conducted at the Center for High-Energy X-ray Sciences (CHEXS), which is supported by the National Science Foundation (BIO, ENG and MPS Directorates) under award DMR-1829070. Part of the research described in this paper was performed at the Canadian Light Source, a national research facility of the University of Saskatchewan, which is supported by the Canada Foundation for Innovation (CFI), the Natural Sciences and Engineering Research Council (NSERC), the National Research Council Canada (NRC), the Canadian Institutes of Health Research (CIHR), the Government of Saskatchewan, and the University of Saskatchewan. Substrate preparation was performed in part at the Cornell NanoScale Facility, a member of the National Nanotechnology Coordinated Infrastructure, which is supported by the NSF (Grant No. NNCI-2025233); the authors would like to thank Sean Palmer and Steven Button for their assistance in substrate preparation and Michel Gingras for discussions.

## Author Contributions

D.G.H. and K.M.S. conceived of, designed, and supervised the experiment. N.K.G. conceived of the experiment and identified anisotropy in the SDW peak shape. The resonant scattering experiments were performed by N.K.G., R.G., M.K., C.T.P., N.C., R.S., and D.G.H. Samples were synthesized and characterized by Y.W. with assistance from C.T.P, D.G.S., and K.M.S. Hard x-ray scattering measurements were conducted and analyzed by B.Z.G., S.S. and A.S. The data were analyzed by N.K.G., R.G., C.T.P., K.M.S. and D.G.H. Model calculations were performed by D.G.H. The manuscript was written by D.G.H. and K.M.S. with input from all authors.

## Competing Interests

The authors declare that they have no competing interests.

## Supplementary Information

Supplementary Information is available for this paper.

# Figures

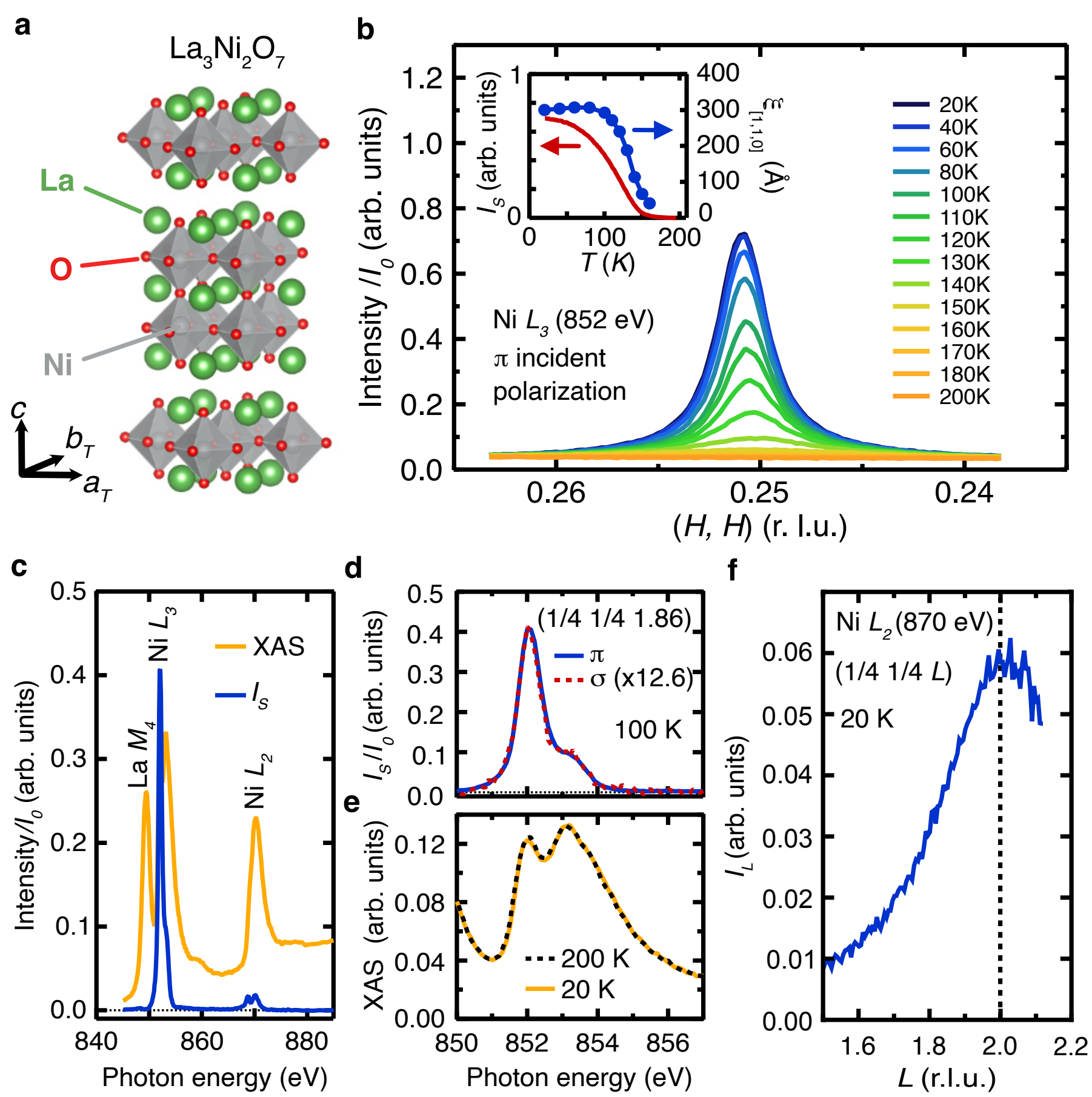

Figure 1: **Magnetic Resonant Scattering and X-Ray Absorption of $La_3Ni_2O_7$.** (**a**) The crystal structure of $La_3Ni_2O_7$, with the pseudo tetragonal axes $a_T$ and $b_T$ denoting the in plane NiO bond directions. (**b**) The intensity (scattering + background) for cuts along the ($H$, $H$, 1.86) direction through the SDW peak at various temperatures, measured at the Ni $L_3$ absorption edge with π incident polarization for sample A. Inset: The SDW peak amplitude, $I_s(T) = I\,(T) - I(220\,K)$, (red) and correlation length along the [1 1 0] direction (blue) as a function of temperature. (**c**) The energy dependence of the SDW peak amplitude, $I_s(\hbar\omega, 100\,K) = I\,(\hbar\omega, 100\,K) - I(\hbar\omega, 220\,K)$, measured with π incident light at $L$ =1.86 through the Ni $L$ and La $M$ edges (blue), along with the x-ray absorption (XAS) measured via partial fluorescence yield (orange). The scattering is peaked at the Ni $L$ resonances but is absent off resonance or at the La $M_4$ resonance. (**d**) The energy dependence of the SDW peak amplitude at 100 K and $L$ = 1.86 at the Ni $L_3$ edge with σ and π incident light. (**e**) The Ni $L_3$ x-ray absorption measured using partial fluorescence yield (PFY) at base temperature (20 K) and above the SDW phase transition (200 K). (**f**) The $L$ dependence of the SDW structure factor, $I_L$, at 20 K of the SDW peak at the Ni $L_2$ edge (870 eV) for sample B.

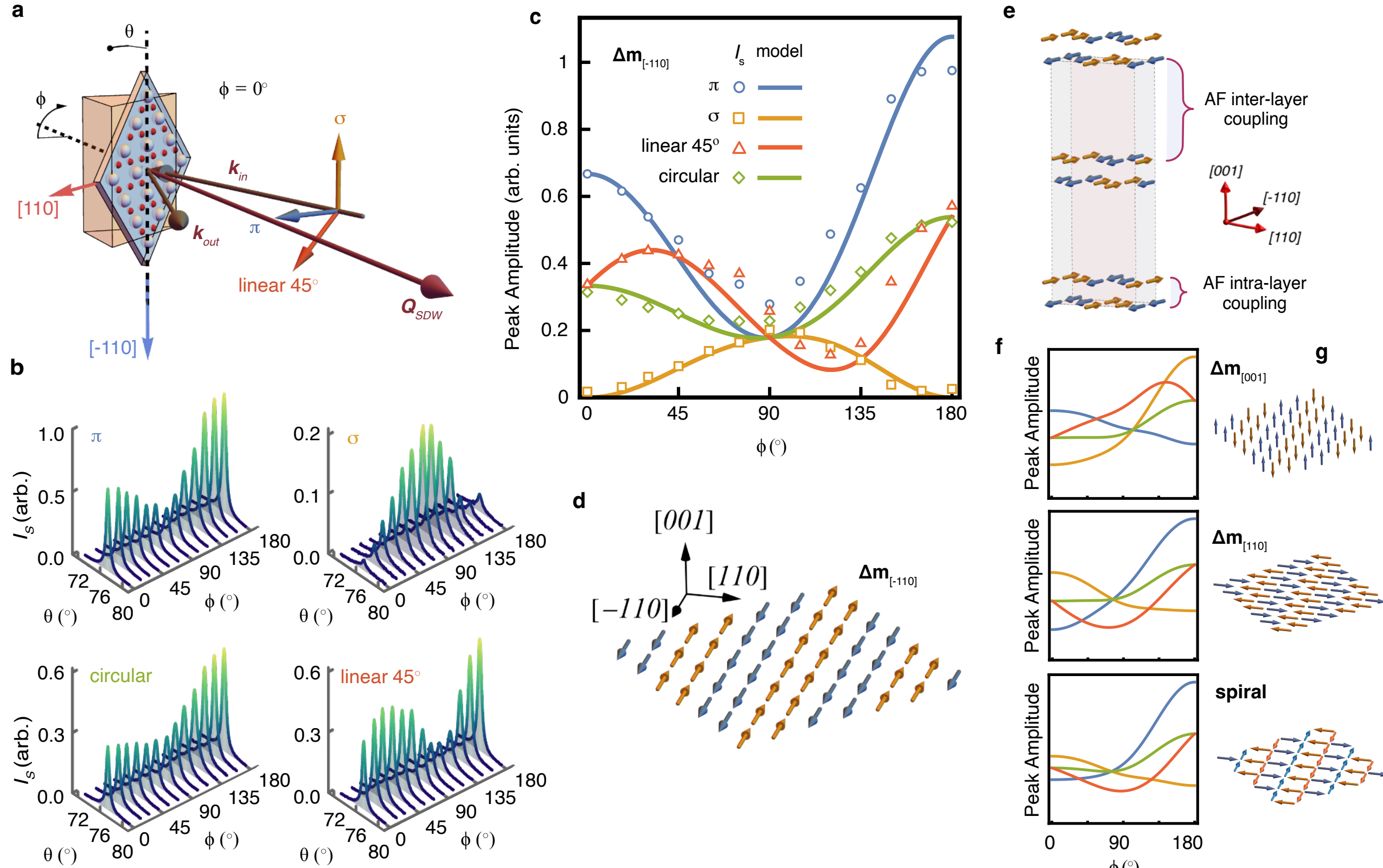

Figure 2: **Measuring Orientation of Magnetic Moments.** (**a**) The measurement geometry for azimuthal angle (ϕ) dependent measurements. The sample is mounted on a 43.7 deg wedge to align the $\left[\frac{1}{4},\frac{1}{4},1.93\right]$ set of planes with the azimuthal rotation axis. The incident photon polarization is set to be σ, π, linear 45° (π − σ) or circular. The sample was rocked about the vertical axis by angle θ to measure the intensity and width of the $\left(\frac{1}{4},\frac{1}{4},1.93\right)$ SDW peak. (**b**) The scattering intensity, $I_s = I\,(T) - I(200\,K)$, versus θ through the SDW peak at ϕ values between 0 and 180 degrees for π, σ, circular and linear 45° incident x-ray polarization at 20 K. The scattering intensity, $I_S$, is found by subtracting the background fluorescence measured above $T_{SDW}$ (200 K) data from the total intensity (scattering + background fluorescence) measured at 20 K. (**c**) The SDW peak amplitude, $I_S$, at 20 K versus ϕ for π, σ, circular and linear 45° incident polarization. Solid lines are the ϕ dependence calculated for a magnetic structure with $\boldsymbol{\Delta m}$ parallel to [−110], depicted in (**d**). (**e**) The 3D magnetic unit cell deduced from the azimuthal angle dependence and assuming anti-ferromagnetic bilayer coupling. (**f**) The ϕ and polarization dependence that would result from the spin configurations depicted in (**g**).

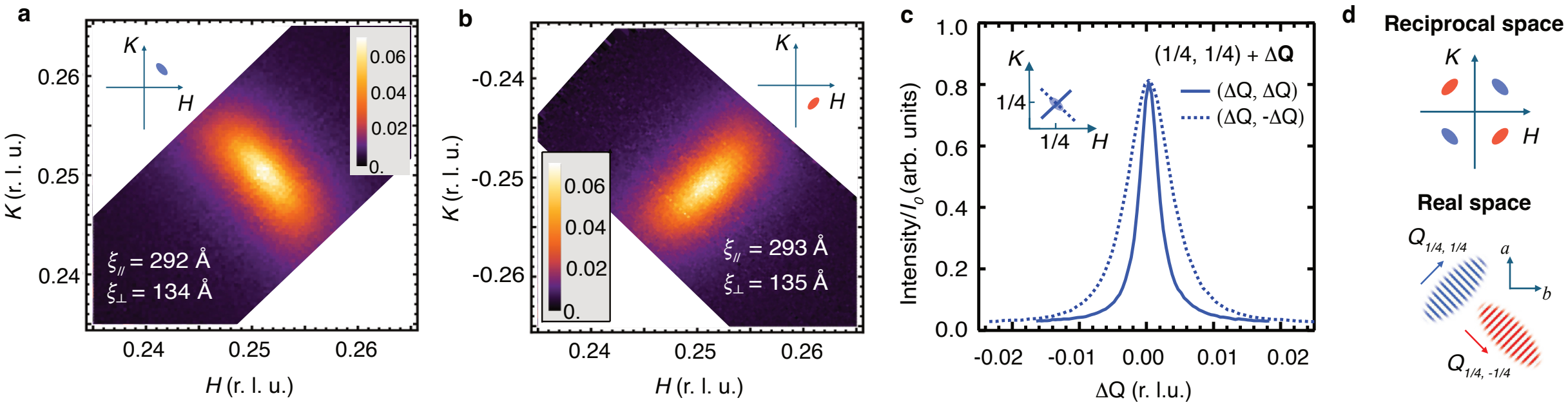

Figure 3: **Anisotropic Bragg peaks indicative of stripe-like SDW domains.** The shape of the (**a**) (1/4 1/4 *L*) and (**b**) (1/4 -1/4 *L*) Bragg peaks at 852 eV with π incident polarization. Measurement of H and K for sample A using a 2D channelplate detector (integrated between $L$ = 1.81 and 1.90). The orthogonal peaks have the same intensities and widths. (**c**) Cuts through the (1/4 1/4 1.85) Bragg peak parallel and perpendicular to the in-plane **Q** vector. (**d**) The shape of Bragg peaks in reciprocal space and the corresponding domain structure in real space for anisotropic domains of unidirectional stripe order, consistent with the measurements in (**a**) and (**b**).

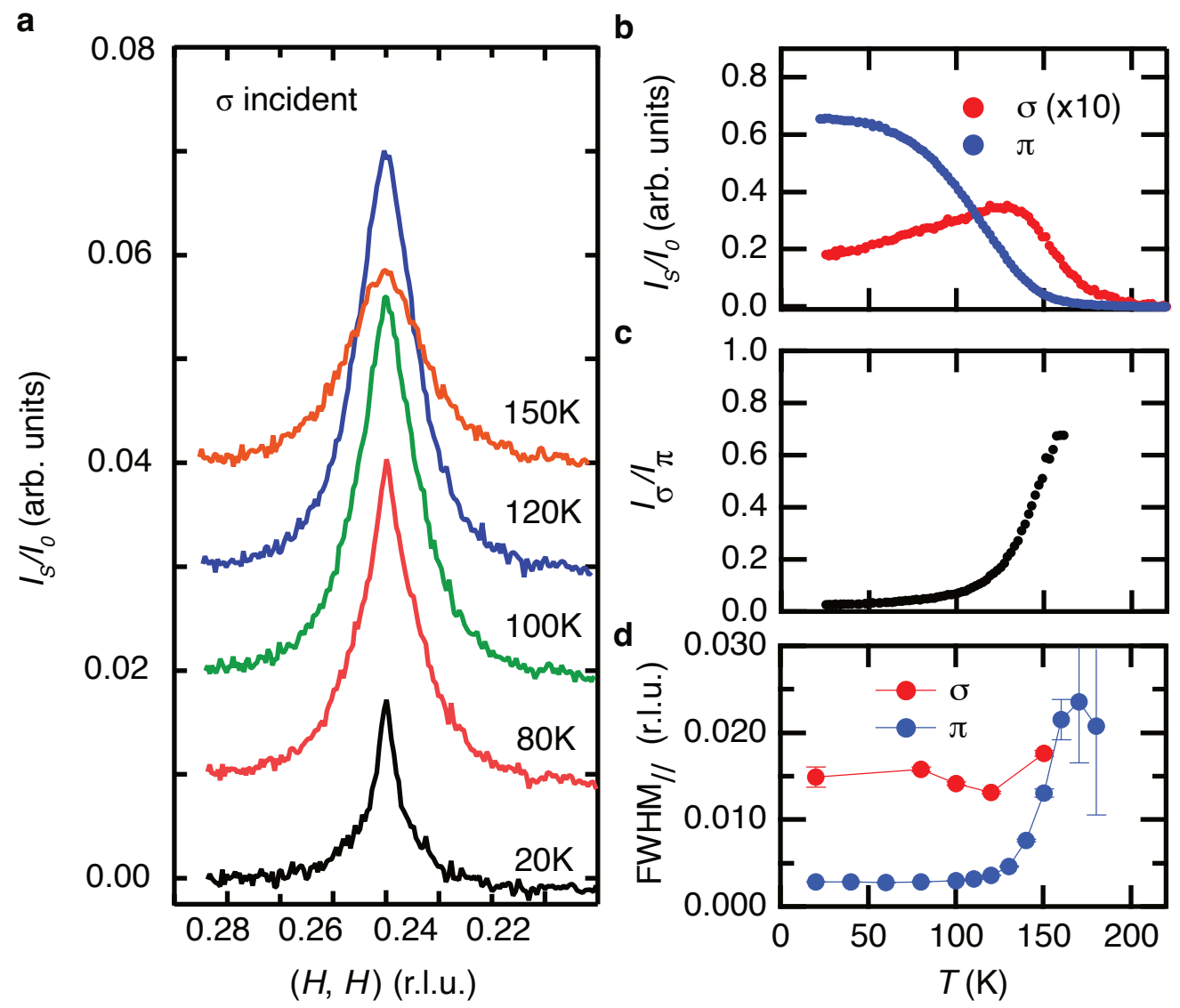

Figure 4. **Temperature dependence of (1/4 1/4 *L*) Bragg peak with σ photon polarization.** (**a**) Cuts of the scattering intensity through the SDW peak at $L$ = 1.93 and 852 eV with incident σ photon polarization at various temperatures. The peak is broader with σ polarization than with π polarization (Fig. 1b) and exhibits a non-monotonic temperature dependence. (**b**) The temperature dependence of the (1/4 1/4 1.93) peak amplitude, $I_s$, with σ and π incident polarization. The peak emerges at 200 K, but peaks at 130 K with σ polarization. (**c**) The ratio of the scattering intensity from σ and π incident light, $I_\sigma/I_\pi$, which increases monotonically with increasing temperature. (**d**) The FWHM of the (1/4 1/4) peak with σ and π incident light. As described in supplementary section VII, $I$s with σ incident light is fit to two peaks (broad and narrow). The narrow peak has FWHM comparable to the $I$s with $\pi$ incident light and is evident at 20 K in a) but diminished above 100 K. We attribute the narrow peak to an artifact of few percent contribution of $\pi$ incident light in the nominal σ polarized beam. The FWHM of the remaining broad peak is shown in d).

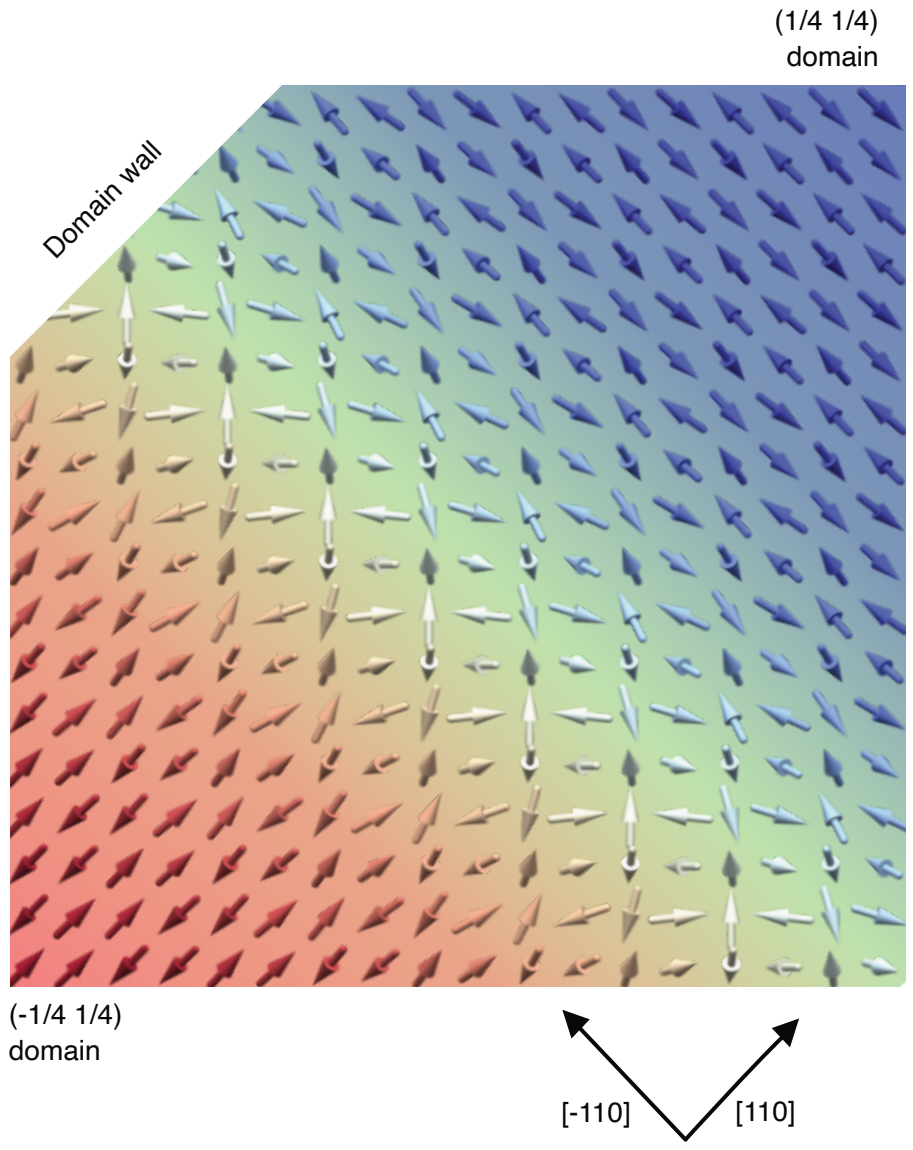

Figure 5: **Conceptual representation of a magnetic domain wall between $\boldsymbol{Q}_{(110)}$ and between $\boldsymbol{Q}_{(-110)}$ SDW domains**. In the vicinity of a domain wall, the spin orientation may rotate to have a sizeable, staggered moment out of the plane and/or parallel to $\boldsymbol{Q}$.